\documentclass[11pt,twoside]{article}
\usepackage{asp2004}
\usepackage{psfig}
\usepackage{epsf}
\usepackage{graphics}
\usepackage{lscape}
\def\edcomment#1{\iffalse\marginpar{\raggedright\sl#1\/}\else\relax\fi}
\reversemarginpar

\begin{document}
\title{Stellar Feedback Processes: Their Impact on Star Formation and Galactic Evolution}

\author{Andreas Burkert}
\affil{University Observatory Munich, Scheinerstr. 1, D-81679 Munich, Germany}

\begin{abstract}
The conditions that lead to self-regulated star formation, star bursts and 
the formation of massive stellar clusters are discussed. 
Massive stars have a strong impact on their environment, especially on 
the evolution of dwarf galaxies which are the building blocks of giant galaxies. 
Energy input by massive young clusters might help to solve some of the most important
puzzles of galaxy formation: the cosmological substructure problem
and the angular momentum problem.
\end{abstract}
\thispagestyle{plain}

\section{Stellar feedback processes and galactic evolution}

Stars and gas in galaxies interact continuously through the exchange of
matter and energy. It is this process that drives galactic evolution and
that determines many of their observed characteristic properties.
Low-mass stars have a limited effect on their environment by
generating wind-blown bubbles and planetary nebulae. High-mass stars, on
the other hand, disturb the dynamics of the surrounding
interstellar medium on larger, that is galactic scales
through their ionizing photons which destroy molecular material
and through supernova explosions which act as point sources of thermal and kinetic energy
input and which drive galactic chemical evolution. The most violent events are 
multiple supernova explosions that erupt on timescales of order $10^6$ yrs 
in young massive clusters (YMCs) and
that can completely change the morphology and evolution of a galaxy by
blowing its gas and metals out into intergalactic space.

\section{Stellar feedback and self-regulated star formation}

Star formation in galaxies is a non-linear process which in details
is not understood up to now (for a recent comprehensive review see
Mac Low  \& Klessen 2004). Ordinary spiral galaxies, like the Milky Way,
"burn" stars at a low, self-regulated rate. Although a large fraction of 
their visible gas is condensed
in cold molecular clouds with masses that by far exceed their thermal Jeans mass, 
star formation turns out to be surprisingly inefficient (Blitz \& Shu 1980).  For example, 
the Milky Way with a total 
molecular gas mass of order $2 \times 10^9 M_{\odot}$ and molecular cloud
densities of order 100 cm$^{-3}$, 
corresponding to collapse timescales of $2 \times 10^6$ yrs, could in principle
form stars with a rate of $10^3$ M$_{\odot}$/yr, which is a factor of 1000 larger than observed.
Irregular supersonic motions (so called "molecular cloud turbulence") have been detected in most
cloud complexes (Larson 1981, Elmegreen \& Falgarone 1996) 
and are considered as the main source for their stability. 
However numerical hydrodynamical simulations consistently show that supersonic, highly compressible turbulence 
dissipates on timescales that are smaller than their collapse timescale
(Stone et al. 1998; Mac Low et al. 1998). In addition, no driver 
of molecular cloud turbulence has ever been found which on the one hand efficiently suppresses
star formation on small scales while, at the same time,
stabilizing giant molecular clouds on the large scales (Heitsch et al. 2001). Magnetic fields 
are not strong enough to stabilize a giant molecular cloud either,
although they might play an important
role on scales of molecular cloud cores, regulating their collapse and angular momentum distribution
and by this also affecting the stellar initial mass function 
and multiplicity (Shu et al. 1987, Mouschovias 1991,
Crutcher 1999).
Kornreich \& Scalo (2000) argued that large-scale interstellar shocks, identified with supernova
remnants or superbubbles, provide an energy source for molecular cloud turbulence. When the shock
passes through a cloud with an internal density gradient, vortical motions are generated that
resemble turbulent motions. Still, it has not yet been demonstrated that such a scenario can
explain the detailed observed structures and velocity fields in molecular clouds.
The virialized, long-lived molecular cloud scenario also leads to a post-T Tauri  problem
(Herbig 1978, Hartmann 2000, Hartmann et al. 2001): the age spread of young stellar populations in 
filamentary cloud complexes like
Taurus are typically 1-3 Myrs which is considerably smaller than their lateral crossing times
that are of order 2-3 $\times 10^7$ yrs. 

These arguments have motivated scientists to explore an alternative scenario,
where molecular clouds never achieve a dynamical equilibrium state of 
turbulent support. They instead form from the turbulent, diffuse interstellar medium with their irregular motions
and clumpy substructures already being imprinted at the time of formation
(Larson 1981, Elmegreen 2000; Hartmann 2003). Star formation then is an 
initial condition problem, rather than determined by the gravitational fragmentation 
of an initially driven, turbulent equilibrium state (Klessen \& Burkert 2000,2001, Bate et al. 2003). 
In this case, the 
important question is not to identify the source of stability and turbulent driving in 
dense clouds, but instead to investigate the various
processes that play a role when molecular clouds form.

Giant molecular clouds are one of the most massive objects in galaxies. How molecular gas
manages to accumulate into such large entities with masses of $10^4 - 10^6 M_{\odot}$ before
condensing into stars and which processes lead to their internal kinematics and density distribution 
is still unclear (Pringle et al. 2001). Hartmann et al. (2001, see also Scalo \& Chappell 1999,
Ballesteros-Paredes et al. 1999, V\'{a}zquez-Semadeni et al. 2003)
suggested that filamentary cloud complexes could form from large-scale 
colliding gas flows (see also Klessen et al, this volume). 
Most of the stellar energy input into the diffuse interstellar medium arises
from supernovae (Wada \& Norman 2001) which drive large-scale turbulent flows with characteristic
velocities of order 10 km/s. Superbubbles, resulting from multiple supernovae
in young massive clusters lead to supersonic large-scale flows with an extent
of 150 pc (Mc Cray \& Kafatos 1987).  When such flows collide, it 
takes of order $10^7$ yrs to accumulate
enough material to form a giant molecular cloud in the post-shock gas. As the age spread of young
stars is a factor of 5 smaller, star formation needs to be suppressed during this stage.
Hartmann et al. (01) argue that this is indeed likely, as the conditions for efficient molecular 
hydrogen formation require a minimum column density of order $1-2 \times 10^{21}$ cm$^{-2}$ for shielding
from the interstellar radiation field to be effective. Prior to that, the dense HI slabs would
still be invisible. 

Figure 1 shows the results of  a numerical calculation of
two converging HI gas flows which collide with Mach 1, leading to a compressed
and cooling dense slab (Burkert, 2004 in preparation). 
A cooling  instability (Burkert \& Lin 2000, Kritsuk \& Norman 2002), combined with a thin slab
instability (Vishniac 1999) enhances small density fluctuations in the incoming HI,
leading to a clumpy, irregular density distribution with internal irregular velocities,
embedded in a hotter environment of compressed HI gas. During the early build-up
phase the internal irregular motions, powered by the kinetic energy of the colliding
gas flows can stabilize the dense 
clumps against gravitational collapse (Koyama \& Inutsuka 2002). After $\sim 10^7$ yrs a
filamentary, highly structured cold cloud complex  has formed that resembles observed
molecular cloud regions well (Fig. 2). Now gravity becomes dominant and
stars begin to form. Within the framework of this scenario, the low rate of star formation in
the Milky Way is a combination of the timescale of molecular cloud formation ($10^7$ yrs), coupled
with a low star formation efficiency of order a few percent.
Note also, that star formation regulates itself as the rate of formation of new sites of star 
formation, as well as their substructures will be determined by
flows that are generated by previous generations of stars.

\begin{figure}
\begin{picture}(0,220)
\put( 75., 0.){\epsfxsize=7.7cm \epsfbox{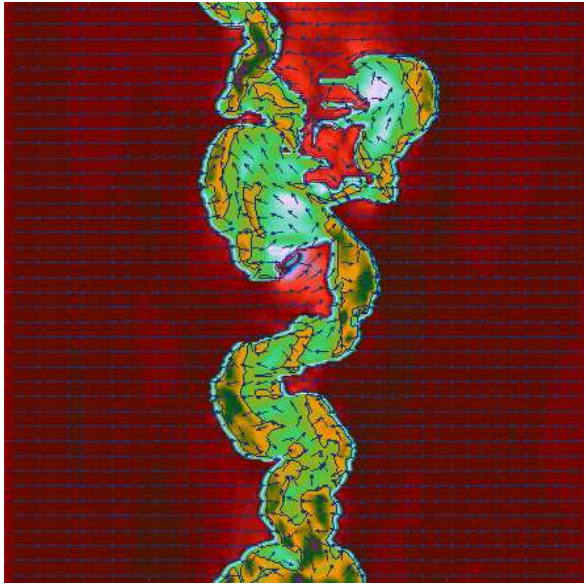}}
\end{picture}
\caption{Result of a numerical 2-dimensional hydrodynamical 
simulation of two colliding gas flows, including cooling.
Details of the irregular velocity field and clumpy substructure
in the dense slab are shown.}
\end{figure}

\begin{figure}
\begin{picture}(0,160)
\put( 40., 0.){\epsfxsize=11.cm \epsfbox{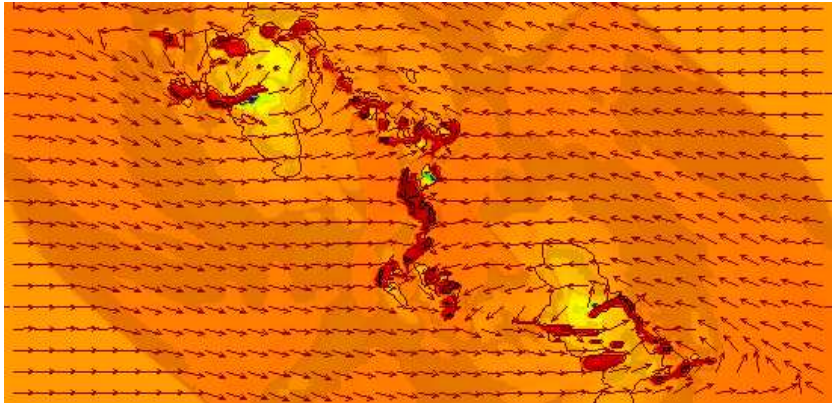}}
\end{picture}
\caption{The slab in a late phase of evolution after $10^7$ yrs.
A system of dense clouds has formed, embedded in a region of compressed HI gas.}
\end{figure}

\section{The origin of star bursts and massive stellar clusters}

Stars tend to form preferentially in clusters (Lada \& Lada 2003, Tan \& McKee, this volume). 
Still, massive clusters with masses exceeding $10^4 M_{\odot}$ are rare in quiescently
evolving, self-regulated galactic environments, like the Milky Way. Larsen (2002) estimates that
only $\sim 15$ \% of normal spirals have clusters as bright as $M_V = -12$. The situation is
different in star burst regions which typically contain swarms of YMCs and super-star clusters that
might even represent the progenitors of globular clusters (Whitmore \& Schweizer 1995, Ashman \& Zepf 1991;
see however Tenorio-Tagle et al. 2003). YMC-formation therefore seems to require unusual conditions
that are found in perturbed galactic environments (Keel \& Borne 2003). 

Interestingly, YMCs follow the same universal mass function $N(m) \sim m^{-2}$ as molecular clouds,
independent of environment. This
can be explained if (a) the probability to form a massive stellar cluster  and
(b) the star formation efficiency, that is the fraction of cloud mass that condenses
into cluster stars, is independent of cloud mass (Elmegreen \& Efremov 1997). As massive clouds are found both
in quiescent as well as in perturbed galaxies, YMCs could in principle form everywhere
and one might at first not expect to find a dependence of the number of YMCs  on environment,
in contradiction with the observations. There exists however strong evidence 
that the star formation efficiency
depends on environment, even if it is independent of cloud mass for a given environment.
Solomon et al. (1997) and Solomon (2000) find that in starburst regions with high star formation efficiencies
(SFE $\approx$ 0.1) the fraction of gas at high densities of $n \geq 10^4 cm^{-3}$ is of 
order 10\%, which is a factor of 10
higher than in unperturbed galaxies like the Milky Way with typical values of SFE $\approx$ 1\%. 
The star formation efficiency therefore seems to be
correlated with the fraction of gas at high densities: 

\begin{equation}
SFE \approx \frac{M_{\geq 10^4 cm^{-3}}}{M_{g}}
\end{equation}

\noindent where M$_{\geq 10^4 cm^{-3}}$ is the total mass of molecular
gas with densities above n = $10^4$ cm$^{-3}$  and M$_g$ is the total cloud mass.
In quiescent regions with SFE $\approx$ 0.01 YMCs
with stellar masses M$_*$ $\geq 10^4$ M$_{\odot}$ can only form
from clouds with masses exceeding M$_g$ $\geq 10^6$ M$_{\odot}$. In perturbed environments with
SFE $\approx 0.1$, on the other hand, 
a much larger fraction of clouds will form YMCs, as the limiting lower mass
for YMC formation is M$_g = 10^5$ M$_{\odot}$.

Again, molecular cloud formation in converging gas flows might be an important mechanism to explain
the large fraction of dense gas in starburst regions as well as their high star formation rates.
In perturbed galactic environments, as for example in merging galaxies,
the relative velocites of colliding gas flows are  $\sim$ 100 km/s,
one to two orders of magnitude larger than in unperturbed disks. Colliding flows
with such a high relative velocity lead to exceptionally high compression 
of the cooling and fragmenting slabs and, by this, to a larger gas fraction at high densities
than in unperturbed galactic environments.
In addition, due to the large inflow velocities, the timescales for accumulating $10^4 - 10^6$ M$_{\odot}$
of gas is reduced by 1-2 orders of magnitudes, leading to 
molecular cloud formation timescales, less than $10^6$ yrs.  Both, the high star formation efficiencies
in combination with the strongly reduced molecular cloud formation timescales could
explain why the star formation rates in perturbed
galacies are orders of magnitudes higher than in unperturbed regions. More theoretical work
is required in order to understand all aspects of this model in details.

\section{Massive clusters and their effect on galaxy evolution}

The  evolution of multiple supernova remnants that form from an OB association
of 10 to 1000 massive stars in YMCs has been summarized by Mac Low (1996).
The most massive stars will become supernovae after just a few million years of the
formation of the YMC. Their winds and supernova blast waves sweep up a
cavity, known as a superbubble, that expands outwards, reaching sizes of order
a few 10 to 100 pc. In the center of the superbubble, the winds and supernova blast waves
of massive stars expand freely until they run into the outer shell, the inner boundary of which is
characterized by a termination shock where the kinetic energy of the freely expanding gas from
the interior is dissipated and converted into thermal energy. The shell 
of gas expands supersonically, sweeping up the 
surrounding interstellar medium in an outer shock. 

In dwarf galaxies with escape velocities of $v_{esc} \leq$ 100 km/s supershells 
will blow out. As the gas pressure scale height in these
systems is smaller than their size, superbubbles accelerate when moving outwards into 
regions of lower pressure until they reach intergalactic space (Mac Low \& McCray 1988). 
De Young \& Heckman (1994) argue that a single YMC with a mass corresponding to a 
1\% burst of star formation could through this mechanism blow out all the baryons 
in spherical galaxies. Detailed hydrodynamical simulations by Mori et al. (2002) also
show that supernova-driven pregalactic 
outflows from subgalactic halos with masses of order $10^8 M_{\odot}$ at redshifts of $z \approx 9$
could efficiently distribute the first products of stellar nucleosynthesis over large cosmological
volumes. 

The blow-away of most baryons in low-mass dark matter halos
by a first generation of YMCs appears to be an attractive solution for
some of the most outstanding puzzles of cosmology: the cosmological substructure problem and the 
angular momentum problem.
As shown in figure 3, cosmological cold-dark-matter simulations predict 10 to 100 times more
satellites around Milky-Way-type galaxies than observed (Moore et al. 1999, Klypin et al. 1999). 
This disagreement could be
accounted for, if most of the low-mass dark halos lost their gas in a first burst of star formation.
Numerical simulations of star formation in
primordial gas clouds in hierarchically forming low-mass dark matter halos suggest that the
initial mass function of the first zero-metallicity stars was extremely top-heavy
(Bromm et al. 1999, 2002; Abel et al. 2000). 
As a result, no visible remnant population would be left behind and most satellites would be
dark and invisible, in agreement with the observations.

\begin{figure}
\begin{picture}(0,140)
\put( 20., 0.){\epsfxsize=12.cm \epsfbox{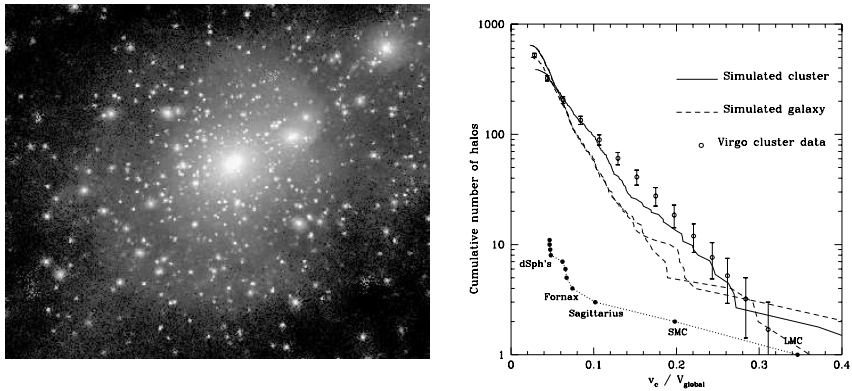}}
\end{picture}
\caption{The left panel shows the dark matter density inside a cosmological
dark matter halo as predicted by numerical simulations
(Ghigna et al. 1998). A large number of satellites is visible.
The right panel, adopted from Moore et al. (1999), shows the
cumulative numbers of dark halos as a function of their circular velocities v$_c$, 
normalized to the circular velocity, v$_{global}$, of the parent halo. Dotted curve: observed
distribution of satellites within the Milky Way. Open circles: observed
distribution of galaxies in the Virgo cluster.
Dashed curves: predicted distribution of satellites in galaxies.
Solid curve: predicted distribution of galaxies in galactic clusters.}
\end{figure}

The standard picture of galactic disk formation assumes that gas cooled and condensed into
low-mass dark matter substructures that subsequently merged to build larger galaxies like
the Milky Way (White \& Rees 1978). During this merging process, collisions of 
substructures occured frequently. While the collisionless dark matter halos could 
move through each
other freely, the collision-dominated gas shocked and was left behind. This process decoupled gas and
dark matter. The gas dissipated its kinetic energy and settled into the equatorial plane, where it formed a 
fast rotating and thin galactic disk.
Detailed hydrodynamical simulations of cosmological disk formation confirm this scenario.
They however also reveal a serious problem: before collisions occur that decouple gas and dark matter,
substructures efficiently lose a large fraction of their specific angular momentum 
by dynamical friction and gravitational torques with the surrounding, diffuse dark matter component and
spiral into the inner regions.
When the gas finally decouples from the dark substructure it has
already lost most of its specific angular momentum, resulting in galactic disks that are an
order of magnitude too small (Steinmetz \& Navarro 1999; Navarro \& Steinmetz 2000).
Again, an early ejection phase triggered by primordial YMCs
could solve this problem (e.g. Robertson et al. 2004).
The metal-enriched gas that had been blown out of the substructures could still be
bound to the larger dark matter density fluctuation that formed the giant galaxy.
If this gas cooled slowly while settling back into the inner regions of the dark halo
without condensing into clumps it would not lose its specific angular momentum
by dynamical friction or tidal torques and could form extended galactic disks, in 
agreement with the observations (Fall \& Efstathiou 1980).

Unfortunately, ejecting most of the gas from a dwarf galaxy is not easy. 
While the superbubbles accelerate outwards, they become Rayleigh-Taylor unstable and fragment.
Their internal high-pressure, hot and metal-loaded gas is
channeled outward through the low-density
gaps in the porous shell (Fujita et al. 2003), while most of the low-metallicity dense gas stays behind.
In addition, galaxies are not spherical. In disks, superbubbles blow out perpendicular
to the disk plane without destroying and ejecting the gaseous disk (Mac Low and Ferrara 1999).

Even if the cosmological substructure problem cannot be solved by
super\-nova-driven galactic winds, most of the metals that are generated by high-mass stars
and released into the hot gas phase by supernovae could still be removed. 
This process might have played an important
role, especially in the early phases of galaxy evolution (redshift $z \approx 10$)
when most of the star formation occured in low-mass dark halos with
masses of order $10^8 M_{\odot}$ (Bromm et al. 2002). Primordial gas infall into these halos and subsequent
fast cooling could have triggered a first central star burst. 
The energy deposited by first-population supernovae could blow away a large fraction of the 
metal-enriched baryons, polluting the surrounding intergalactic medium
(Mori et al. 1997, Bromm et al. 2003) with metals. This scenario would explain the 
origin of the heavy elements that have been
detected in the Ly $\alpha$ forest clouds at $z \approx 3$. During the galactic
wind phase some metals were mixed into the dense gas that was left behind in the galaxy. 
A second generation of metal-poor stars formed from this gas component.
The stellar populations of those
substructures that later on merged with larger galaxies would today populate the their
halos.  Dwarf galaxies and satellites of giant galaxies represent those low-mass substructures
that survived disruption by merging with larger galaxies. Their strikingly narrow
metallicity-luminosity correlations (Skillman et al. 1989) could 
provide interesting insight into the early star formation phases of galaxies
and the energetic processes that dominated galaxy formation at that time.


\begin{thebibliography}{} 
\bibitem[Abel et al. (2000)]{ab2000}
Abel, T., Bryan, G.L. \& Norman, M.L. 2000, ApJ, 540, 39
\bibitem[Ashman \& Zepf (2001)]{az2001}
Ashman, K.M. \& Zepf, S.E. 2001, AJ, 122, 1988
\bibitem[Ballesteros-Paredes et al. (1999)]{bh1999}
Ballesteros-Paredes, J., Hartmann, L. \& V\'{a}zquez-Semadeni, E. 1999, ApJ, 527, 285
\bibitem[Blitz \& Shu (1980)]{bs1980}
Blitz, L. \& Shu, F.H. 1980, ApJ, 238, 148
\bibitem[Bromm et al. (1999)]{bc1999}
Bromm, V., Coppi, P.S. \& Larson, R.B. 1999, ApJ, 527, L5
\bibitem[Bromm et al. (2002)]{bc2002}
Bromm, V., Coppi, P.S. \& Larson, R.B. 2002, ApJ, 647, 23
\bibitem[Bromm et al. (2003)]{by2003}
Bromm, V., Yoshida, N. \& Hernquist, L. 2003, ApJ, 596, L135
\bibitem[Burkert \& Lin (2000)]{bl2000}
Burkert, A. \& Lin, D.N.C. 2000, ApJ, 537, 270
\bibitem[Crutcher (1999)]{cr1999}
Crutcher, R.N. 1999, ApJ, 520, 706
\bibitem[de Young \& Heckman (1994)]{dh1994}
de Young, D.S. \& Heckman, T.M. 1994, ApJ, 431, 598
\bibitem[Elmegreen \& Efremov (1997)]{ee1997}
Elmegreen, B.G. \& Efremov, Y. 1997, ApJ, 480, 235
\bibitem[Elmegreen \& Falgarone (1996)]{ef1996}
Elmegreen, B.G. \& Falgarone, E. 1996, ApJ, 471, 816
\bibitem[Elmegreen (2000)]{el2000}
Elmegreen, B.G. 2000, ApJ, 530, 277
\bibitem[Fall \& Efstathiou (1980)]{fe1980}
Fall, S.M. \& Efsthatiou, G. 1980, MNRAS, 193, 189 
\bibitem[Fujita et al. (2003)]{fm2003}
Fujita, A., Martin, C.L., Mac Low, M.-M. \& Abel, T. 2003, ApJ, 599, 50
\bibitem[Hartmann (2000)]{ha2000}
Hartmann, L. 2000, in 33d ESOLAB Symp., Star Formation from the
Small to the Large Scale, ed. F. Favata, A. Kaas \& A. Wilson (ESA SP-445), p107
\bibitem[Hartmann et al (2001)]{ha2001}
Hartmann, L., Ballesteros-Paredes, J \& Bergin, E.A. 2001, ApJ, 562, 852
\bibitem[Hartmann (2003)]{ha2003}
Hartmann, L. 2003, ApJ, 585, 398
\bibitem[Heitsch et al (2001)]{hm2001}
Heitsch, F., Mac Low, M.-M. \& Klessen, R.S. 2001, ApJ, 547, 280
\bibitem[Herbig (1978)]{h1978}
Herbig, G.H. 1978, in Problems of Physics and Evolution of the
Universe, ed. L.V. Mirzoyan (Yerevan: Acad. Sci. Armenian SSR), p171
\bibitem[Keel \& Borne (2003)]{kb2003}
Keel, W.C. \& Borne, K.D. 2003, ApJ, 126, L257
\bibitem[Klypin et al. (1999)]{kl1999}
Klypin, A., Kravtsov, A.V., Valenzuela, O. \& Prada, F. 1999, ApJ, 522, 82
\bibitem[Kornreich \& Scalo (2000)]{ks2000}
Kornreich, P. \& Scalo, J. 2000, ApJ, 531, 366
\bibitem[Koyama \& Inutsuka (2002)]{ki2002}
Koyama, H. \& Inutsuka, S. 2002, ApJ, 564, L97
\bibitem[Kritsuk \& Norman (2002)]{kn2002}
Kritsuk, A.G. \& Norman, M.L. 2002, ApJ, 569, L127
\bibitem[Lada \& Lada (2003)]{ll2003}
Lada, C.J. \& Lada, E.A. 2003, ARA\&A, 41, 57
\bibitem[Larsen (2002)]{la2002}
Larsen, S. 2002, AJ, 124, 1393
\bibitem[Larson (1981)]{la1981}
Larson, R.B. 1981, MNRAS, 194, 809
\bibitem[Mac Low \& McCray (1988)]{mm1988}
Mac Low , M.-M. \& McCray, R. 1988, ApJ, 324, 776
\bibitem[Mac Low (1996)]{ml1996}
Mac Low, M.-M. 1996, in The Interplay between Massive Star Formation, the ISM
and Galaxy Evolution, eds. D. Kunth, B. Guiderdoni, M. Heydari-Malayeri \& T.X. Thuan,
(Editions Frontiers), p. 169
\bibitem[Mac Low et al. (1998)]{mk1998}
Mac Low, M.-M., Klessen, R.S., Burkert, A. \& Smith, M.D. 1998, Phys. Rev. Lett., 80, 2754
\bibitem[Mac Low \& Ferrara (1999)]{mf1999}
Mac Low, M.-M. \& Ferrara, A. 1999, ApJ, 513, 142
\bibitem[Mac Low \& Klessen (2004)]{mk2004}
Mac Low, M. \& Klessen, R.S. 2004, Reviews of Modern Physics, 76, 125
\bibitem[McCray \& Kafatos (1987]{mk1987}
McCray, R. \& Kafatos, M. 1987, ApJ, 317, 190
\bibitem[Moore et al. (1999]{mo1999}
Moore, B., Ghigna, S., Governato, F., Lake, G., Quinn, T. \& Stadel, J. 1999,
ApJ, 524, L19
\bibitem[Mori et al (1997]{mo1997}
Mori, M., Yoshii, Y., Tsujimoto, T. \& Nomoto, K. 1997, ApJ, 478, L21
\bibitem[Mori et al (2002]{mo2002}
Mori, M., Ferrara, A. \& Madau, P. 2002, ApJ, 571, 40
\bibitem[Mouschovias (1991]{mu1991}
Mouschovias, T.C. 1991, ApJ, 373, 169
\bibitem[Navarro \& Steinmetz (2000]{ns2000}
Navarro, J.F. \& Steinmetz, M. 2000, ApJ, 538, 477
\bibitem[Pringle et al (2001]{pa2001}
Pringle, J.E., Allen, R.J. \& Lubow, S.H. 2001, MNRAS, 327, 663
\bibitem[Robertson et al (2004]{ry2004}
Robertson, B., Yoshida, N., Springel, V. \& Hernquist, L. 2004, astro-ph/0401252
\bibitem[Scale \& Chappell (1999)]{sc1999}
Scalo, J. \& Chappell, D. 1999, MNRAS, 310, 1
\bibitem[Shu et al. (1987)]{sa1987}
Shu, F.H., Adams, F.C. \& Lizano, S. 1987, ARAA, 25, 23
\bibitem[Skillman et al. (1989)]{sk1989}
Skillman, E.D., Kennicutt, R.C. \& Hodge, P. 1989, ApJ, 347, 875
\bibitem[Solomon et al. (1997)]{sd1997}
Solomon, P.M., Downes, D., Radford, S.J.E. \& Barrett, J.W. 1997, ApJ, 478, 144
\bibitem[Solomon (2000)]{so2000}
Solomon, P.M. 2000, in Starburst Galaxies: Near and Far, eds. L. Tacconi \& D. Lutz,
(Springer, Berlin), p. 173
\bibitem[Steinmetz \& Navarro (1999)]{sn1999}
Steinmetz, M. \& Navarro, J.F. 1999, ApJ, 513, 555
\bibitem[Stone et al. (1998)]{so1998}
Stone, J.M., Ostriker, E.C. \& Gammie, C.F. 1998, ApJ, 508, L99
\bibitem[Tenorio-Tagle et al. (2003)]{tp2003}
Tenorio-Tagle, G., Palous, J., Silich, S., Medina-Tanco, G.A. \& Casiana, M.T.
2004, AA, 411, 397
\bibitem[Vazquez-Semadeni et al.(2003)]{va2003}
V\'{a}zquez-Semadeni, E., Ballesteros-Paredes, J. \& Klessen, R.S. 2003, ApJ, 585, L131
\bibitem[Vishniac (1994)]{v1994}
Vishniac, E.T. 1994, ApJ, 428, 186
\bibitem[Wada \& Norman (2001)]{wn2001}
Wada, K. \& Norman, C.A. 2001, ApJ, 547, 172
\bibitem[White \& Rees (1978)]{wr1978}
White, S.D.M. \& Rees, M.J. 1978, MNRAS, 183, 341
\bibitem[Whitmore \& Schweizer (1995)]{ws1995}
Whitmore, B.C. \& Schweizer, F. 1995, AJ, 109, 960
\end{thebibliography}
\end{document}